\documentstyle[a4,12pt]{article}

\newcommand{\as}{\mbox{$\alpha_s$}}
\newcommand{\go}{\mbox{$\tilde g$}}

\newcommand{\lgl}{light gluino}
\newcommand{\dis}{deep inelastic scattering}

\newcommand{\sm}{standard model}

\newcommand{\rg}{renormalization group}

\newcommand{\susic}{supersymmetric}

\newcommand{\trm}{transverse momentum}

\newcommand{\cm}{center of mass}

\newcommand{\no}{neutralino}

\newcommand{\ep}{\mbox{$e^+e^-$}}

\def\lr3{$SU(3)_L\otimes SU(3)_R$}

\def\z0{$Z^0$}
\def\Z0{$Z^0$}

\def\ep{$e^+e^-$}

\def\nlo{next-to-leading order}
\def\nnlo{next-to-next-to-leading order}

\def\cm{centre of mass}
\def\gsim{\buildrel{\lower.7ex\hbox{$>$}}\over{\lower.7ex\hbox{$\sim$}}}
\def\lsim{\buildrel{\lower.7ex\hbox{$<$}}\over{\lower.7ex\hbox{$\sim$}}}
\newenvironment{comment}[1]{}{}

\begin{document}

\thispagestyle{empty}
\setcounter{page}{0}

\begin{flushright}
LMU-09/93\\
August 1993
\end{flushright}
\vspace*{\fill}

\begin{center}
{\LARGE\bf Hadron Multiplicities in the Presence of Light Gluinos}\\
\vspace{2cm}
{\Large Frank Cuypers}\footnote{\small
	Email: {\tt frank@hep.physik.uni-muenchen.de}}\\
{\it Sektion Physik der Universit\"at M\"unchen, 80333 M\"unchen, FRG}
\end{center}
\vspace*{\fill}

\centerline{\large Abstract}
\begin{quote}
We consider the effect of a massless gluino
on the evolution of a parton shower.
The hadron multiplicity distribution
is predicted to evolve as a function of the collider energy
in the same way as in the absence of gluinos,
except for a slower running of $\alpha_s$.
A comparison of the predicted average hadron multiplicity
with experimental data
is presented as a fit
constraining the values
of the strong coupling constant at the $Z^0$ peak $\alpha_s(m_Z)$
and the scale of the onset of hadronization $m_0$.
It appears that \lgl s
are unlikely to be detected this way
with less than 500 GeV
of \cm\ energy.
\end{quote}
\vspace*{\fill}

\newpage
\section{Introduction}

It has recently been emphasized \cite{LG2,LG3}
that present experimental data
still allow for the existence of a \lgl,
with a mass typically of the order of a few GeV.
At first sight,
this might seem paradoxical:
how could a light,
strongly interacting particle
have remained undetected up to now?

The reason
is that within the framework of the MSSM
a gluino can only be produced alongside with another \susic\ particle,
a squark or another gluino.
If this spartner is a squark,
detection at the present accelerator facilities
is not possible
because of the high mass of the squarks.
However,
if two gluinos are produced from a parent gluon,
the kinematical limit does not present a problem
and one would expect an abundant gluino pair production
in hadronic reactions.
Unfortunately,
the splitting kernel of a gluon into two gluinos
is not singular,
so that gluino pair production is a \nlo\ effect\footnote{The situation
is analogous for the effects of the pair production
of one quark-antiquark flavour,
which is suppressed by
a factor $1/2N_c^3$, where $N_c$ is the number of colours.}.

Still,
one might hope that some \dis\ or jet observables
are sufficiently influenced
by the existence of \lgl s,
for deviations from pure QCD
to be measurable.

It is the purpose of this paper
to compute the impact of massless gluinos
on a particularly simple jet characteristic,
namely the average hadron multiplity.
For this we introduce in the next section
a simple model of hadronization,
which has the virtue of depending only on one parameter.
We then compute the way a parton shower is modified
in the presence of massless gluinos.
The resulting evolution equations
are then solved analytically
and the solution is compared to data.

\section{Hadronization}

In the following we shall assume
that gluinos hadronize as ordinary partons.
This is not an unreasonable assumption,
since they will emerge inside a jet
as a blanched hadronic bound state,
probably a gluino-gluon system.
This so-called ``glueballino''
will eventually decay into two light quarks and an invisible \no.

Within the framework of the {\em Lund string fragmentation model} \cite{PS14}
and of {\em local parton-hadron duality} \cite{PS15},
we assume that the multiplicity distribution of hadrons $P_n$
emerging at the end of a parton cascade
is obtained from the distribution $P(\lambda)$
of the {\em string length} $\lambda$
by the convolution
\begin{equation}
P_n = \int_0^\infty d\lambda~K(n,\lambda)~P(\lambda)
\label{e0}
{}~.
\end{equation}
Note that $P_n$ is a discrete distribution
while $P(\lambda)$ is continuous
and can be determined from perturbative QCD.
The non-perturbative effects of QCD
are contained in the hadronization kernel $K$,
which is assumed to have the following properties:
\begin{itemize}
\item	it is local $K(n,\lambda) = K(n-\lambda)$;
\item	it is symmetric $K(x) = K(-x)$;
\item	it does not depend on the energy of the process $K \ne f(\sqrt{s})$.
\end{itemize}
Dirac or Gaussian distributions,
{\em e.g.},
naturally satisfy these requirements,
but the exact shape of the hadronization kernel
cannot be predicted by perturbative QCD.
\begin{comment}{
$K(x) = \delta(x)$
or
$K(x) = \sqrt{2\over\pi\sigma^2} e^{-\left({n-\lambda\over\sigma}\right)^2}$,
}\end{comment}
However,
if the energy of the process is sufficiently large,
{\em i.e.} if the logarithm of the energy scale is much larger
than the width of the kernel
$L = \ln{s/\Lambda^2} \gg \sigma_K$
the average $\overline{n}$ and the width $\sigma_n^2$
of the hadron multiplicity distribution
are given by
\begin{eqnarray}
\overline{n} &=& \overline{\lambda}
\label{e-2}
\\
\sigma_n^2 &=& \sigma_\lambda^2 + \sigma_K^2
\label{e-3}
{}~,
\end{eqnarray}
where $\overline{\lambda}$ and $\sigma_\lambda$
are the mean and the width of the string length distribution.
It is important to notice
that the average hadron multiplicity
does not depend on the hadronization kernel.
Since the string length distribution
can be computed from perturbative QCD
if the energy scale $m_0$
at which the parton cascade stops
and hadronization occurs is given,
this scale is thus the only hadronization parameter
which is needed to compute the average hadron multiplicity.

In principle,
a further parameter is needed to describe hadronization.
This is a proportionality constant
between the average hadron multiplicity
and the average string length
in Eq.~(\ref{e-2}).
This constant takes different values
according to
whether charged or neutral hadrons
and whether heavy flavour decays
are taken into account.
Still,
this parameter is very much correlated to $m_0$.
As it turns out
it can safely be set to one,
which we do in what follows.
The fit to data
presented in the last section
could only be marginally improved by another choice.

Furthermore,
since the hadronization kernel does not depend on the energy of the process
whereas $\sigma_\lambda$ increases rapidly with the jet energy,
at sufficiently high energies
hadronization will not influence very much
the shape of the multiplicity distribution.
The cross-over energy
for which $\sigma_\lambda>\sigma_K$
is around 50 GeV \cite{PS13}.

\section{Parton Shower}

There exist several formalisms
to describe a parton cascade \cite{PS6,PS5,PS12,PS2}.
Since they are all based on the {\em modified leading log approximation}
\cite{PS5}
they are all expected to be yield the same final results.
We choose here to adopt the method described in Ref.~\cite{PS12},
because of its elegant underlying geometrical picture
and because it allows a straightforward implementation
of the important \nnlo\ recoil effects \cite{PS3,PS18}.

To compute the distribution $P(\lambda)$
of the string length
in the presence of massless gluinos,
we closely follow the method presented in Ref.~\cite{PS12}.
{}From this distribution and Eq.~(\ref{e-2})
we should extract the average particle multiplicity
and compare with experimental data.

Gluinos manifest themselves
\begin{description}
\item[indirectly,]
	by modifying the running of the strong coupling constant \as;
\item[directly,]
	by the splitting of a virtual gluon into two gluinos
	and the subsequent radiation of gluons
	off this colour dipole.
\end{description}

The value of \as\ at a given scale $Q^2$ is dictated by the \rg\ equation
\begin{equation}
\as = {1 \over \beta\ln {Q^2\over\Lambda^2}}
{}~,
\label{e1}
\end{equation}
where the $\beta$ function depends on the relevant group theory factors $T$
of the $SU(3)$ representation,
{\em i.e.} the adjoint for the gluon $g$ and the gluino $\go$
and the fundamental for the quark $q$:
\begin{equation}
T(g)=T(\go)=T({\rm adjoint})= N_c
\qquad\qquad
T(q)=T({\rm fundamental})= {1\over2}
{}~.
\label{e3}
\end{equation}
The $\beta$ function is thus given to first order by
\begin{equation}
\beta = {1 \over 4\pi}
\left[
	{11\over3} T(g) - N_f {4\over3} T(q) - {2\over3} T(\go)
\right]
= {1 \over 12\pi}
\left[
	9 N_c - 2 N_f
\right]
{}~,
\label{e2}
\end{equation}
where $N_c$ and $N_f$ are the number of colours and active flavours
respectively.
The factor $-2/3$ in the gluino term
is due to its Majorana fermion nature.

In view of Eq.~(\ref{e2})
it is sometimes claimed that the effect of light gluinos,
at the parton level,
is very closely approximated by increasing $N_f$,
the number of active quark flavours,
by 3.
This claim is not always correct.
Indeed,
in a parton shower,
massless gluinos are also produced as real particles
and participate in two ways to a QCD cascade:
\begin{enumerate}
\item	The splitting of a gluon into a gluino pair,
	as depicted in Fig.~\ref{f1}.
	This effect can indeed be mimicked by setting $N_f\to N_f+3$.
\item	The radiation of a gluon off a gluino pair,
	as depicted in Fig.~\ref{f2}.
	This effect cannot be mimicked by setting $N_f\to N_f+3$.
\end{enumerate}

The distribution for a gluon to split into a pair of gluinos
of \trm\ $k_\perp$ and energy fractions $z$ and $1-z$
is given by
\begin{equation}
dn =
N_c~{\alpha_s\over2\pi}~d\kappa~dz~P_{\tilde gg}
{}~,
\label{e100}
\end{equation}
where $\kappa = \ln{k_\perp^2 / \Lambda^2}$
and the splitting function
$P_{\tilde gg} = {1\over2}\left[ z^2+(1-z)^2 \right]$
is non-singular.
It is convenient to define the (true) constant $\alpha_0$
\begin{equation}
{\alpha_0 \over \kappa}
=
N_c~{\alpha_s\over2\pi}
\qquad\Rightarrow\qquad
\alpha_0 = {6N_c\over9N_c-2N_f}
{}~,
\label{e103}
\end{equation}
in terms of which we define
the gluino pair production constant $\alpha_{p\tilde g}$
\begin{equation}
{\alpha_{p\tilde g} \over \kappa}
=
N_c~{\alpha_s\over2\pi}~\int_0^1 dz~P_{\tilde gg}
\qquad\Rightarrow\qquad
\alpha_{p\tilde g} = {1\over3}\alpha_0
{}~.
\label{e101}
\end{equation}

Once a gluino pair has been produced
it will itself contribute to the parton cascade
by radiating more gluons
according to the distribution
\begin{eqnarray}
dn
&=&
{\alpha_0\over\kappa}~d\kappa~dy~
\left[
	1 - 2e^{\kappa-L\over2}\cosh y + e^{\kappa-L}\cosh 2y
\right]
\end{eqnarray}
which can be approximated by a rapidity plateau
\begin{eqnarray}
dn
&\approx&
{\alpha_0\over\kappa}~d\kappa~dy~~
\theta(L-\kappa-c_q-2|y|)
\qquad (L\gg\kappa)
{}~,
\label{e104}
\end{eqnarray}
where $c_q=3/2$,
$L=\ln s/\Lambda^2$
and $y$ is the rapidity of the emitted gluon.

It is straightforward to incorporate these processes
into the formalism of Ref.~\cite{PS12}.
The Laplace transform of the string length distribution
\begin{equation}
{\cal P}^{(i)}(\beta)
=
\int_0^\infty d\lambda~P^{(i)}(\lambda)~e^{-\beta\lambda}
\qquad\qquad (i=q,g,\go)
{}~,
\label{e8}
\end{equation}
obeys then the coupled retarded differential energy evolution equations
{\samepage
\begin{eqnarray}
{d^2\over dL^2} \ln{\cal P}^{(g)}
&=&
{\alpha_0 \over (L-c_g)}
\left[ {\cal P}^{(g)}(L-c_g)-1 \right]
\label{e401}\\
&&
+
{d\over dL}
\Biggl\{
{\alpha_{pq} \over L}
\left[
	  \ln{\cal P}^{(q)}(L-{c_p\over2})
	+ \ln{\cal P}^{(g)}(L-{c_p\over2})
	-2\ln{\cal P}^{(g)}(L)
\right]
\nonumber\\
&&
\qquad
+
{\alpha_{p\tilde g} \over L}
\left[
	  \ln{\cal P}^{(\tilde g)}(L-{c_p\over2})
	+ \ln{\cal P}^{(g)}(L-{c_p\over2})
	-2\ln{\cal P}^{(g)}(L)
\right]
\Biggr\}
\nonumber\\
{d^2\over dL^2} \ln{\cal P}^{(q)}
&=&
{\alpha_q \over (L-c_q)}
\left[ {\cal P}^{(g)}(L-c_q)-1 \right]
\label{e402}
\\
{d^2\over dL^2} \ln{\cal P}^{(\tilde g)}
&=&
{\alpha_0 \over (L-c_q)}
\left[ {\cal P}^{(g)}(L-c_q)-1 \right]
\label{e403}
\end{eqnarray}
}
\begin{comment}{

\begin{eqnarray}
{d\over dL} \ln{\cal P}^{(g)}
&=&
R^{(g)}(L)
+
{\alpha_{pq} \over L}
\left[
	  \ln{\cal P}^{(q)}(L-{c_p\over2})
	+ \ln{\cal P}^{(g)}(L-{c_p\over2})
	-2\ln{\cal P}^{(g)}(L)
\right]
\label{e401}
\\
&&\qquad\qquad
+
{\alpha_{p\tilde g} \over L}
\left[
	  \ln{\cal P}^{(\tilde g)}(L-{c_p\over2})
	+ \ln{\cal P}^{(g)}(L-{c_p\over2})
	-2\ln{\cal P}^{(g)}(L)
\right]
\nonumber\\
{d\over dL} \ln{\cal P}^{(q)}
&=&
R^{(q)}(L)
\label{e402}
\\
{d\over dL} \ln{\cal P}^{(\tilde g)}
&=&
R^{(\tilde g)}(L)
\label{e403}
\\
{d\over dL} R^{(g)}
\quad&=&
{\alpha_0 \over (L-c_g)}
\left[ {\cal P}^g(L-c_g)-1 \right]
\label{e404}
\\
{d\over dL} R^{(q)}
\quad&=&
{\alpha_q \over (L-c_q)}
\left[ {\cal P}^g(L-c_q)-1 \right]
\label{e405}
\\
{d\over dL} R^{(\tilde g)}
\quad&=&
{\alpha_0 \over (L-c_q)}
\left[ {\cal P}^g(L-c_q)-1 \right]
\label{e406}
\end{eqnarray}

}\end{comment}
where
\begin{equation}
c_q = {3\over2}
\qquad\qquad
c_g = {11\over6}
\qquad\qquad
c_p = {13\over6}
\label{e6}
\end{equation}
\begin{equation}
\alpha_{q} = \alpha_0 \left( 1-{1\over N_c^2} \right)
\qquad\qquad
\alpha_{pq} = {1\over3} \alpha_0 {N_f\over N_c}
\qquad\qquad
\alpha_{p\tilde g} = {1\over3} \alpha_0
{}~.
\label{e7}
\end{equation}

These equations have to be solved with the boundary conditions
\begin{equation}
{\cal P}^{(i)}(\beta;L=\kappa_c) = 1
\qquad\qquad
{d\over dL}{\cal P}^{(i)}(\beta;L=\kappa_c) = -\beta
\label{e9}
{}~,
\end{equation}
where
\begin{equation}
\kappa_c = \ln {m_0^2 \over \Lambda^2}
\label{e91}
\end{equation}
and $m_0$ is the scale of the onset of hadronization.

With these boundary conditions
(and the fact that only the limit $\beta=0$ is of any interest,
as will become clear shortly)
it is clear that the gluino degrees of freedom can be incorporated into the
quark ones.
Indeed,
Eqs~(\ref{e402},\ref{e403}) combine to
\begin{equation}
\ln{\cal P}^{(\tilde g)} = {\alpha_0\over\alpha_q} \ln{\cal P}^{(q)}
\label{e10}
{}~.
\end{equation}
\begin{comment}{

Indeed,
Eqs~(\ref{e405},\ref{e406}) and Eqs~(\ref{e402},\ref{e403}) combine to
\begin{equation}
R^{(\tilde g)} = {\alpha_0\over\alpha_q} R^{(q)}
\qquad\qquad
\ln{\cal P}^{(\tilde g)} = {\alpha_0\over\alpha_q} \ln{\cal P}^{(q)}
\label{e10}
{}~.
\end{equation}

}\end{comment}
We can thus rewrite the full system \ref{e2}-\ref{e7}
in terms of two second order equations
\begin{eqnarray}
{d^2\over dL^2} \ln{\cal P}^{(g)}
&=&
{\alpha_0 \over (L-c_g)}
\left[ {\cal P}^{(g)}(L-c_g)-1 \right]
\label{e11}
\\
&&
+
{d\over dL}
\left\{
	 {\alpha' \over L} \ln{\cal P}^{(q)}(L-{c_p\over2})
	+{\alpha''\over L}
	 \left[
		+ \ln{\cal P}^{(g)}(L-{c_p\over2})
		-2\ln{\cal P}^{(g)}(L)
	 \right]
\right\}
\nonumber
\\
{d^2\over dL^2} \ln{\cal P}^{(q)}
&=&
{\alpha_q \over (L-c_q)}
\left[ {\cal P}^{(g)}(L-c_q)-1 \right]
\label{e12}
{}~,
\end{eqnarray}
where we defined
\begin{equation}
\alpha' = {\alpha_0\over3} \left( {N_f\over N_c}+{\alpha_0\over\alpha_q}
\right)
\qquad {\rm and} \qquad
\alpha'' = {\alpha_0\over3} \left( {N_f\over N_c}+1 \right)
\label{e13}
{}~.
\end{equation}

Expanding the functions ${\cal P}$ and $\ln{\cal P}$
in powers of $\beta$,
we obtain
\begin{eqnarray}
{\cal P}(L;\beta)
&=&
1 + \sum_{l=1}^\infty{(-\beta)^l\over l!}M_l(L)
\label{e14}
\\
\ln{\cal P}(L;\beta)
&=&
\qquad
\sum_{l=1}^\infty{(-\beta)^l\over l!}H_l(L)
\label{e15}
{}~,
\end{eqnarray}
where the coefficients $M_l$ are the moments
of the string length distribution $P(\lambda)$
\begin{equation}
M_l = \overline{\lambda^l}
\label{e16}
\end{equation}
and the coefficients $H_l$ are given by
{\samepage
\begin{eqnarray}
H_1 &=& M_1
\nonumber\\
H_2 &=& M_2-M_1^2
\nonumber\\
H_3 &=& M_3-3M_1M_2+2M_1^3
\label{e17}\\
H_4 &=& M_4-4M_1M_3-3M_2^2+12M_1^2M_2-6M_1^4
\nonumber\\
H_5 &=& M_5-5M_1M_4-10M_2M_3+20M_1^2M_3
\nonumber\\
&&-60M_1^3M_2+30M_1M_2^2+24M_1^5
\nonumber\\
\vdots
\nonumber
\end{eqnarray}
}
Each power of $\beta$
yields a set of two coupled equations,
which recursively determine the moments of $P(\lambda)$ \cite{PS16}:
\begin{eqnarray}
{d^2\over dL^2} H_l^{(g)}
&=&
\alpha_0
{M_l^{(g)}(L-c_g) \over (L-c_g)}
\label{e18}
\\
&&
+
\alpha' {d\over dL}
\left[
	{H_l^{(g)}(L-{c_p\over2}) \over L}
\right]
\nonumber
\\
&&
+
\alpha'' {d\over dL}
\left[
	{H_l^{(g)}(L-{c_p\over2}) \over L} - 2{H_l^{(g)}(L) \over L}
\right]
\nonumber
\\
{d^2\over dL^2} H_l^{(q)}
&=&
\alpha_q
{M_l^{(g)}(L-c_q) \over (L-c_q)}
\label{e19}
{}~,
\end{eqnarray}
with the boundary conditions
\begin{equation}
\begin{array}{cccccl}
\left.M_l\right|_{L=\kappa_c}
&=& \left.H_l\right|_{L=\kappa_c} &=& 0
\\
\displaystyle
\left.{\partial M_l\over\partial L}\right|_{L=\kappa_c}
&=&\displaystyle \left.{\partial H_l\over\partial L}\right|_{L=\kappa_c} &=& 0
&\qquad (l\ge2)
\\
\displaystyle
\left.{\partial M_1\over\partial L}\right|_{L=\kappa_c}
&=&\displaystyle \left.{\partial H_1\over\partial L}\right|_{L=\kappa_c} &=& 1
\label{bc}
\end{array}
\end{equation}

\section{Analytic Solution}

These equations (\ref{e18}-\ref{bc})
can easily be solved analitically
if we make two approximations:
\begin{enumerate}
\item	since the last two terms of Eq.~(\ref{e18}) are of higher order,
	we can neglect them for the determination of $H_l^{(q)}$
	and assume
	\begin{equation}
	H_l^{(q)} = {\alpha_q\over\alpha_0} H_l^{(g)}
	\label{e20}
	~;
	\end{equation}
\item	since at sufficiently high energies $L\gg c_q,c_g$,
	the retardation can be approximated by a Taylor expansion
	around $L$.
\end{enumerate}
The coupled retarded differential equations (\ref{e18},\ref{e19})
reduce then to the single differential equation
\begin{equation}
{H_l^{(g)}}'' = \alpha_0
\left[
	  {M_l^{(g)} \over L}
	- c_g \left( {M_l^{(g)} \over L} \right)'
	- {N_f\over3N_c^3} \left( {H_l^{(g)} \over L} \right)'
\right]
\label{e21}
{}~.
\end{equation}
Except for the slower running of $\alpha_s$
(which is taken into account by $\alpha_0$ in Eq.~(\ref{e103}))
this is nothing else than the result which would have been obtained
in the absence of gluinos!
This comes as a disappointment,
since this way of hunting \lgl s
turns thus out to be equivalent
to just an intricated measurement of the running of $\alpha_s$.

The cancellation of the gluino contribution
can be explained in the following way.
In the large $N_c$ limit
gluons behave as quark-antiquark pairs
and the splitting of a gluon into a real quark-antiquark pair
goes almost unnoticed.
This implies a strong colour suppresion
of the order of $N_f/N_c^3$.
In the case of gluinos
there is no need to invoke the large $N_c$ limit,
since for the colour flow
the splitting of a gluon into a gluino pair
is identical with the splitting of a gluon into two gluons:
instead of a colour suppression
there is an exact cancellation
(at least with the approximations we have made,
which hide the effects of the fermionic nature of the gluino).

In virtue of Eq.~(\ref{e-2}),
the first multiplicity moment
does not depend on the details of hadronization
and can be compared directly with experimental data.
For this purpose
we write the solution to Eq.~(\ref{e21}) for $l=1$
with the boundary conditions (\ref{bc})
in terms of modified Bessel functions \cite{PS12a}.
For the average hadron multiplicity in an $e^+e^-$ event
we find thus
\begin{eqnarray}
\overline{n}
{}~=~
\overline{\lambda^{(q)}}
&=& {\alpha_q\over\alpha_0} ~\overline{\lambda^{(g)}}
\nonumber
\\
&=& \left( 1-{1\over N_c^2} \right)
  {zz_0\over2\alpha_0}
  \left( {z_0\over z} \right)^B
  \left[
	K_{1+B}(z_0) I_{1+B}(z) - I_{1+B}(z_0) K_{1+B}(z)
  \right]
\label{e22}
{}~,
\end{eqnarray}
where
\begin{equation}
\alpha_0={6N_c \over 9N_c-2N_f}
\qquad
B={11+2{N_f\over{N_c}^3}\over 9-2{N_f\over N_c}}
\label{e23}
\end{equation}
\begin{equation}
z=\sqrt{4\alpha_0L}
\qquad
z_0=\sqrt{4\alpha_0\kappa_c}
\label{e24}
\end{equation}
\begin{equation}
L = \ln {s\over\Lambda^2}
\qquad
\kappa_c = \ln {m_0^2\over\Lambda^2}
\label{e25}
{}~.
\end{equation}
Note that for pure QCD
the only modification to these equations
consists of replacing Eqs (\ref{e23}) by
\begin{equation}
\alpha_0={6N_c \over 11N_c-2N_f}
\qquad
B={11+2{N_f\over{N_c}^3}\over 11-2{N_f\over N_c}}
{}~.
\label{e26}
\end{equation}
The average hadron multiplicity,
as given by Eq.~(\ref{e22}),
depends thus on two parameters:
\begin{enumerate}
\item	The hadronization scale $m_0$,
	{\em i.e.} the energy scale at which the parton cascade stops
	and hadronization starts.
	It is of the order of a few hadron masses,
	but cannot be computed from perturbative QCD
	and constitutes thus the main obstacle to an accurate prediction.
\item	The (one loop, no threshold) QCD scale $\Lambda$,
	as it is defined in Eq.~(\ref{e1}).
	We present our results in the next section in terms of the value of
	the strong coupling constant at the $Z^0$ peak
	\begin{equation}
	\alpha_s(m_Z) = \alpha_0 {2\pi\over N_c}{1\over\ln{m_Z^2\over\Lambda^2}}
	\label{e27}
	~,
	\end{equation}
	which is tightly constrained by existing experimental data.
\end{enumerate}

\section{Comparison with Data}

A $\chi^2$ fit of Eq.~(\ref{e22})
(with $N_f=5$)
to the data of Refs~\cite{PS17}
yields the following values for the best fit:
\begin{eqnarray}
{\rm without~gluino:} && \alpha_s(m_Z)=.114 \qquad m_0=.321~{\rm GeV} \qquad
\chi^2/{\rm dof}=.434
\label{e28}
\\
{\rm with~gluino:} && \alpha_s(m_Z)=.126 \qquad m_0=.291~{\rm GeV} \qquad
\chi^2/{\rm dof}=.412~.
\label{e29}
\end{eqnarray}
The errors corresponding to the desired confidence level
can be deduced from the $\chi^2$ contours in Fig.~\ref{f3}.
Note that the central values of $\alpha_s(m_Z)$ (\ref{e28},\ref{e29})
almost coincide with previous determinations \cite{LG1}
extracted from jets data at LEP and low energy data
in the absence and the presence of \lgl s.
This provides a powerful consistency check.
The strong correlation between the two parameters
should come as no surprise,
in view of the very similar role
$\Lambda$ and $m_0$ play in Eq.~(\ref{e22}).
The relatively low values obtained for the hadronization scale $m_0$
are due to the fact that the approximate (analytic) solution (\ref{e22})
to Eqs~(\ref{e18},\ref{e19})
slightly underestimates the exact (numerical) solution \cite{PS12a}.

Anyhow,
one should not grant the values obtained for this parameter
too much significance,
since the data from which it is fitted
contain only the charged tracks
and include strange particle decay products.
As we already mentioned,
this effect should be parametrized
by a proportionality constant between $\overline{n}$ and $\overline{\lambda}$.
Since any improvement of the fits (\ref{e28},\ref{e29})
would not add any physical significance,
we have kept the value of this extra parameter equal to one.
The addition of extra data points at other energies
(there are many more available)
would not change anything to the conclusions.

The low $\chi^2$ values of the two fits (\ref{e28},\ref{e29})
are demonstrated in Fig.~\ref{f4},
where we have plotted the energy dependence of the average hadron multiplicity
according to Eq.~(\ref{e22})
with (\ref{e23})
and without (\ref{e26}) massless gluinos
alongside with the experimental data of Refs~\cite{PS17}.
Clearly,
the present experimental data cannot distinguish
between pure QCD
and SUSY with a massless gluino
(or {\em a fortiori} a \lgl).
It is only starting from a collider energy of approximately 500 GeV
that the difference between the predictions
for the average hadron multiplicities in $e^+e^-$ collisions
becomes of the same order as the experimental error.
For higher multiplicity moments
the situation is even worse,
since hadronization affects the predictions more (\ref{e-2}).

\section{Conclusion}

We have computed the incidence of a \lgl\ on the evolution of a parton shower
and estimated the resulting effects on the average hadron multiplicity in jets.
For this we have used a formalism
which takes into account the most important \nnlo\ effects
within the framework of the {\em modified leading log approximation}.
Since the observable we analyze is a very inclusive quantity,
we have argued that it is reasonnable to assume that
a gluino hadronizes in the same way as ordinary partons
and that hadronization can be modeled here with only one parameter.

To summarize our results,
the hadron multiplicity distribution in jets
is very little affected by the presence of \lgl s.
In the approximation we use to solve analytically
the jet evolution equations,
the effect of gluinos actually cancels out totally,
except for a slower running of $\alpha_s$.
We conclude that measuring deviations from the \sm\ prediction
of the hadron multiplicity distribution in jets
is not likely to shed any light on the \lgl\ puzzle
at the present \ep\ colliders or at LEP2.

\bigskip

I am indebted to G\"osta Gustafson
for having familiarized me with his elegant calculation scheme
and for having kept me abreast of all developments
prior to publication.
I am also thankful to Geert Jan van Oldenborgh
for suggesting this calculation
and to Marcela Carena, Alexander Khodjamirian, Wolfgang Ochs and Reinhold
R\"uckl
for useful discussions.

\newpage

\section{Figures}

\begin{figure}[htb]
\begin{center}
\begin{picture}(300,200)(0,-50)
\Gluon(0,50)(25,50){3}{4}
\ArrowLine(25,50)(75,100)
\ArrowLine(75,0)(25,50)
\Text(37.5,-15)[tc]{$dn = N_f~{\alpha_s\over2\pi}~d\kappa~dz$}
\Text(37.5,-35)[tc]{${1\over2}\left[ z^2+(1-z)^2 \right]$}
\Gluon(175,50)(200,50){3}{4}
\Line(200,50)(250,100)
\Line(250,0)(200,50)
\Text(212.5,-15)[tc]{$dn = N_c~{\alpha_s\over2\pi}~d\kappa~dz$}
\Text(212.5,-35)[tc]{${1\over2}\left[ z^2+(1-z)^2 \right]$}
\end{picture}
\end{center}
\caption[dummy]{Splitting of a gluon into two partons.}
\label{f1}
\end{figure}

\begin{figure}[htb]
\begin{center}
\begin{picture}(350,210)(0,-60)
\Vertex(0,50){2}
\Gluon(0,50)(25,75){3}{5}
\Gluon(25,75)(50,100){3}{5}
\Gluon(0,50)(50,0){3}{10}
\Gluon(25,75)(50,75){3}{4}
\Text(25,-15)[tc]{$dn = N_c~{\alpha_s\over4\pi}~dx_1~dx_2$}
\Text(25,-35)[tc]{$\displaystyle{x_1^3+x_2^3 \over (1-x_1)(1-x_2)}$}
\Vertex(150,50){2}
\ArrowLine(150,50)(175,75)
\ArrowLine(175,75)(200,100)
\ArrowLine(200,0)(150,50)
\Gluon(175,75)(200,75){3}{4}
\Text(175,-15)[tc]{$dn = {N_c\over2}\left(1-{1\over
N_c^2}\right)~{\alpha_s\over2\pi}~dx_1~dx_2$}
\Text(175,-35)[tc]{$\displaystyle{x_1^2+x_2^2 \over (1-x_1)(1-x_2)}$}
\Vertex(300,50){2}
\Line(300,50)(325,75)
\Line(325,75)(350,100)
\Line(300,50)(350,0)
\Gluon(325,75)(350,75){3}{4}
\Text(325,-15)[tc]{$dn = N_c~{\alpha_s\over4\pi}~dx_1~dx_2$}
\Text(325,-35)[tc]{$\displaystyle{x_1^2+x_2^2 \over (1-x_1)(1-x_2)}$}
\end{picture}
\end{center}
\caption[dummy]{Radiation of a gluon off parton pairs.
	The energy fractions of the recoiled parent partons
	are denoted $x_1$ and $x_2$.}
\label{f2}
\end{figure}

\begin{figure}[htb]
\begin{center}
\begin{picture}(554,504)(60,0)
\put(0,0){\strut\epsffile{/user/frank/Gluino/MultAnal/contour.ps}}
\Text(80,450)[tr]{\Large $m_0$ [GeV]}
\Text(481,19)[tr]{\Large $\alpha_s(m_Z)$}
\end{picture}
\end{center}
\caption[dummy]{Contours of $\Delta\chi^2=1,2,3$
for the fit of the average hadron multiplicity
in the plane of the two free parameters
$\alpha(m_Z)$ and $m_0$.}
\label{f3}
\end{figure}

\begin{figure}[htb]
\begin{center}
\begin{picture}(554,504)(60,0)
\put(0,0){\strut\epsffile{/user/frank/Gluino/MultAnal/fit.1000.ps}}
\put(160,30){\strut\epsffile{/user/frank/Gluino/MultAnal/fit.100.ps}}
\Text(60,450)[tr]{\Large $\overline{n}$}
\Text(481,19)[tr]{\Large $\sqrt{s}$ [GeV]}
\end{picture}
\end{center}
\caption[dummy]{Best fit of the average hadron multiplicity
as a function of the collider energy.}
\label{f4}
\end{figure}

\end{document}